# Conditioning noise is a free regularizer for LoRA fine-tuning: no pathology encoder required for diffusion-based artifact detection in histopathology

Konstantinos Moutselos (corresponding author; kmouts@unipi.gr; ORCID 0000-0002-6759-8540) and Ilias Maglogiannis (imaglo@unipi.gr; ORCID 0000-0003-2860-399X) — Department of Digital Systems, University of Piraeus, Piraeus, Greece.

## Abstract

Diffusion-based artifact detectors score whole-slide image patches by reconstruction error under a model fine-tuned on clean tissue. We show that conditioning this fine-tuning on random Gaussian embeddings — resampled at every step from ~200 KB of precomputed embedding statistics, with no encoder, no cache, and no change to inference — consistently widens the clean/artifact separation. A four-step ablation chain shows the benefit requires neither content (shuffled real embeddings), provenance (synthetic Gaussians), a tuned intensity (flat across an 8× variance range), nor per-patch identity (fresh per-step noise); a LoRA-dropout control shows the conditioning pathway specifically, not generic weight perturbation, carries the effect. Patch-level gains of +0.25–0.48 Cohen's d replicate across nine trainings; honest leave-one-slide-out evaluation clears a pre-registered bar in 2/2 seeds; and two pre-registered external endpoints on a 281-case set confirm pooled ΔF1 = +0.0073 (95% CI) and +0.0129 (97.5% CI, two-look corrected). We release the full evaluation protocol, including measured seed noise and selection-optimism pricing.

## 1. Introduction

Whole-slide images enter computational pipelines carrying the accidents of their making: out-of-focus regions, air bubbles, tissue folds, pen markings, foreign debris. Undetected, these artifacts corrupt downstream analysis — silently, since most models score artifacts as confidently as tissue. Automated quality control is therefore a gating step for computational pathology at scale, and its hardest requirement is coverage without annotation: artifact taxonomies are open-ended, and pixel-level ground truth is expensive where it exists at all.

Reconstruction-based detection answers the annotation problem by modeling only clean tissue: a generative model fine-tuned on artifact- free patches reconstructs them well and artifacts poorly, and the error becomes the detector. Diffusion models have made this paradigm practical, but it inherits a structural weakness — the better the backbone generalizes, the better it reconstructs what it should reject. A detector built on a strong foundation model must be *prevented* from extending its competence to artifacts; regularization, not capacity, is the scarce resource.

This paper contributes a regularizer that costs nothing. During LoRA fine-tuning of a latent diffusion backbone, we feed the cross-attention conditioning pathway fresh Gaussian noise at every step — noise shaped only by ~200 KB of precomputed embedding statistics. Inference is byte-identical to the unconditioned baseline. The method emerged from a finding we did not anticipate: conditioning the fine-tuning on real pathology-encoder embeddings helps detection, but *why* it helps has nothing to do with what the embeddings contain. A four-way ablation chain (Section 4) shows the benefit survives shuffling the embeddings across patches, replacing them with moment-matched synthetic Gaussians, scaling their variance across an 8x range, and abandoning fixed per-patch embeddings for fresh per-step draws — while a LoRA-dropout control shows generic weight noise does not reproduce it. The foundation encoder, the original point of the design, turns out to be the one removable part.

We validate under a protocol built for effects near the noise floor: measured seed noise at patch and slide level, leave-one-slide-out threshold selection with the common shortcut's optimism priced explicitly, and two pre-registered external endpoints on a 281-slide external set — both confirmed, the second under a declared two-look multiplicity correction ($\Delta$F1 +0.0073 at 95%; +0.0129 at 97.5%). A feature-density baseline family answers "why diffusion" and reveals an inverse specialization between reconstruction- and feature-based detection that we believe is itself useful to the field.

Contributions: (i) a training-only, annotation-free, foundation-model-free regularizer for diffusion-based artifact detection with zero inference cost; (ii) a mechanism characterization by elimination across twelve trainings; (iii) a pre-registered, noise-calibrated evaluation protocol with two confirmed external endpoints; (iv) findings about what does *not* work or does not transfer — including a data-scaling regime that worsens detection and an artifact class whose detectability collapses across sites — reported with the same rigor as the positive results.

## 2. Related work

### 2.1 Artifact detection and QC in computational pathology

Supervised QC pipelines detect known artifact classes with dedicated segmentation models trained on annotated corpora [1,2]. Their coverage is bounded by their label sets, and pixel-level ground truth is scarce or absent for several clinically common artifact types (no public dataset we surveyed provides pen-marking masks). Reconstruction-based detection inverts the data requirement: model clean tissue only, and flag what reconstructs poorly. DiffusionQC [3] instantiates this with a latent diffusion backbone fine-tuned on clean patches and scored by denoising error; our work builds directly on this paradigm (Section 3.1) and modifies only its training.

### 2.2 Diffusion models for unsupervised anomaly detection

Diffusion-based AD spans medical imaging [4] and industrial inspection [5]. Two convergent observations from recent, unrelated domains support our design choices: mid-range timesteps emerge as the informative scoring window in IC parametric-test screening as in our $t$ = 650/800 setting [6] — and partial-chain scoring was the operative observation of the earliest diffusion-AD work [4] — and reformulations that correct "deviations from normality" in latent space [7] pursue the same identity-shortcut problem our regularizer addresses from the training side. Adaptive per-sample denoising depth [5] is complementary to, and orthogonal from, our training-only intervention.

### 2.3 Zero-shot and feature-density alternatives

Vision-language ZSAD adapts CLIP-family models to name anomalies in unseen domains [8]; its "zero-shot" transfers anomaly knowledge from auxiliary *annotated* corpora, whereas our setting assumes no anomaly annotation exists anywhere in the pipeline. Feature-density methods (kNN/PaDiM/PatchCore lineage [9,10]; GMM over filtered features [11]) model the normal feature distribution directly. We evaluate this family head-to-head in Section 6.4 and find it inversely specialized relative to reconstruction error — strong on structurally foreign content, weak on photometric degradation — and dependent on the foundation encoder at inference, the dependency our method removes.

### 2.4 Noise injection, conditioning, and regularization

Training with input noise is classically equivalent, to first order, to Jacobian-norm regularization [12]; our mechanism section reads the conditioning-noise effect in exactly this family, with the ablation chain excluding content-, identity-, and intensity-based accounts. Conditioning dropout in classifier-free guidance [13] intermittently *removes* conditioning to shape generation; we *always replace* it with fresh noise, at training only, for a discrimination objective — kin in mechanism, different in target. Random pairing of conditions has also served data-free distillation of generative models [14] — the same mechanical act, aimed at condition-space coverage rather than regularization, with conditioning retained at inference. Weight-space stochasticity over adapters (masked LoRA experts [15]; contrastive adapter geometry during and after training [16,17]) perturbs a different pathway; our dropout control (Section 4.6) indicates pathway identity matters, marking this family as a boundary rather than an equivalent.

### 2.5 Positioning

Against this landscape the contribution is narrow and, we believe, sharp: a training-only, annotation-free, foundation-model-free regularizer for reconstruction-based artifact detection, characterized by elimination (what it does not need) and confirmed under a pre-registered two-look external protocol. We are not aware of prior work employing conditioning-pathway noise purely as a train-time regularizer for anomaly scoring.

## 3. Method

### 3.1 Base pipeline

We build on the reconstruction-error paradigm of DiffusionQC, in our independent re-implementation (the reproduction and its evaluation are the subject of a companion study [18]; here the pipeline is fixed infrastructure). Patches of 1024x1024 px at 10x are encoded by the frozen VAE of the PixCell-1024 latent-diffusion backbone [19]; a LoRA adapter (rank 16, alpha 16; the 224 attention projections; 8.26M trainable parameters) is fine-tuned for 1,000 steps (batch 4, lr 1e-5, ~2 h on one A100) on 2,185 clean patches from 16 slides, using the standard epsilon-prediction objective. At inference a patch is scored by the mean epsilon-prediction error at a fixed timestep under the *null* (empty) conditioning embedding; whole-slide heatmaps tile this score at stride 1024 over tissue. Throughout, "encoder-free" refers to the removal of the pathology foundation model (UNI2-h) from all stages; the latent VAE of the PixCell backbone is retained, as in all reconstruction-based scoring.

**Metrics.** Patch-level separation is reported as Cohen's d between the artifact and clean score distributions of a held-out validation pool (1,167 clean and 260 artifact patches from two withheld slides): the difference of group means divided by the pooled standard deviation, positive when artifacts score higher; AUROC over the same pool accompanies it where informative. Slide-level and external performance are reported as F1 of binarized artifact maps against ground truth, pooled over all evaluation units (cells or pixels) rather than averaged per slide, so that large and small slides contribute in proportion to their area.

### 3.2 Data

Development data derive from the 50 TCGA whole-slide images (H&E, scored at 10x) carrying public artifact annotations from AIRAQc [20], as used by the DiffusionQC benchmark [3]. AIRAQc defines eight artifact classes but releases evaluated annotations for four — air-bubble, pen-marking, out-of-focus, and fold — which constitute the ground-truth taxonomy here (our companion study [18] documents this accounting, including a decoding defect in an earlier five-class reading of the palette). Eight of the 50 slides are excluded as data properties recorded with the split (2 unusable, 6 low-tissue); the remaining 42 follow the partition fixed in the companion study and inherited unchanged — 16 training, 2 validation, 24 test — so no split decision was made with knowledge of this paper's results. **Training:** 2,185 clean 1024-px patches from the 16 training slides, selected by a tissue-fraction gate and curated artifact-exclusion criteria; the same slides contribute 400 artifact patches (natural mix: air-bubble 47%, pen-marking 29%, out-of-focus 18%, fold 5%) used only by contrastive arms not part of the proposed method. **Validation pool (patch-level diagnostics):** 1,167 clean and 260 artifact patches from the 2 held-out slides. **Slide-level test set:** the 24 test slides with AIRAQc pixel ground truth, disjoint from all slides above. **External set:** the GrandQC MPP10 artifact test cohort [1] — 281 cases (one whole-slide image per case) across four organs with technician-induced, expert-annotated artifacts under GrandQC's own class taxonomy (33,361 evaluated 2x2-tile blocks; 134.4M valid cells per model) — used exclusively under the frozen two-look protocol of Section 5.3, with no role in any development choice.

### 3.3 The proposed modification

The entire proposal is one change to training, and none to inference. Figure 1 summarizes all three stages. Offline, once, the first and second moments of UNI2-h embeddings are estimated over the clean training patches, separately for each position of the 16-token conditioning sequence (one token per 256-px subpatch of the 4 x 4 grid, 1,536 dimensions each — the backbone's native conditioning interface):

```
mu, sd in R^{16 x 1536},  ~200 KB.
```

During fine-tuning, at every step and for every patch in the batch, the cross-attention conditioning input receives a fresh draw

```
e = mu + sd * eps,   eps ~ N(0, I_{16 x 1536}),
```

with elementwise multiplication, from a fixed-seed generator. That is the complete recipe: no encoder call, no embedding cache, no architectural change, no new loss term, no schedule. Scoring remains the unconditioned error of Section 3.1 — byte-identical to the baseline detector — so the method's inference cost, latency, and memory are exactly those of the model it improves.

### 3.4 Cost accounting

Training overhead is one Gaussian sample per step (negligible against the transformer forward/backward); the moment estimation is a single offline pass over the clean set, and only its 200 KB summary ships with the training code. Against the self-conditioning alternative that motivated this line, the proposal removes: the 681M-parameter UNI2-h encoder from the training loop, the 205 MiB embedding cache, and every version/provenance concern attached to a foundation-model dependency. Against the unconditioned baseline, it adds nothing at deployment — which is why we describe the mechanism as a *free* regularizer: its entire price is paid, in kilobytes, before training begins.

### 3.5 Pre-registration and seeds

Every experimental claim in this paper was pre-registered in a public, version-controlled lab record before the corresponding run, with frozen reading rules (effect bars in units of measured seed noise) and, for external claims, frozen thresholds and a declared multiplicity correction. Training-seed noise — the standard deviation of a model's final metric across trainings that differ only in random seed — was measured from three baseline replicates before any comparison was interpreted: $s_{seed} = 0.126$ in Cohen's d at patch level, $s_{slide} = 0.0112$ in pooled F1 at slide level; every effect bar in this paper is a pre-declared multiple of these floors. Section 5 details the protocol; the record itself accompanies the code release.

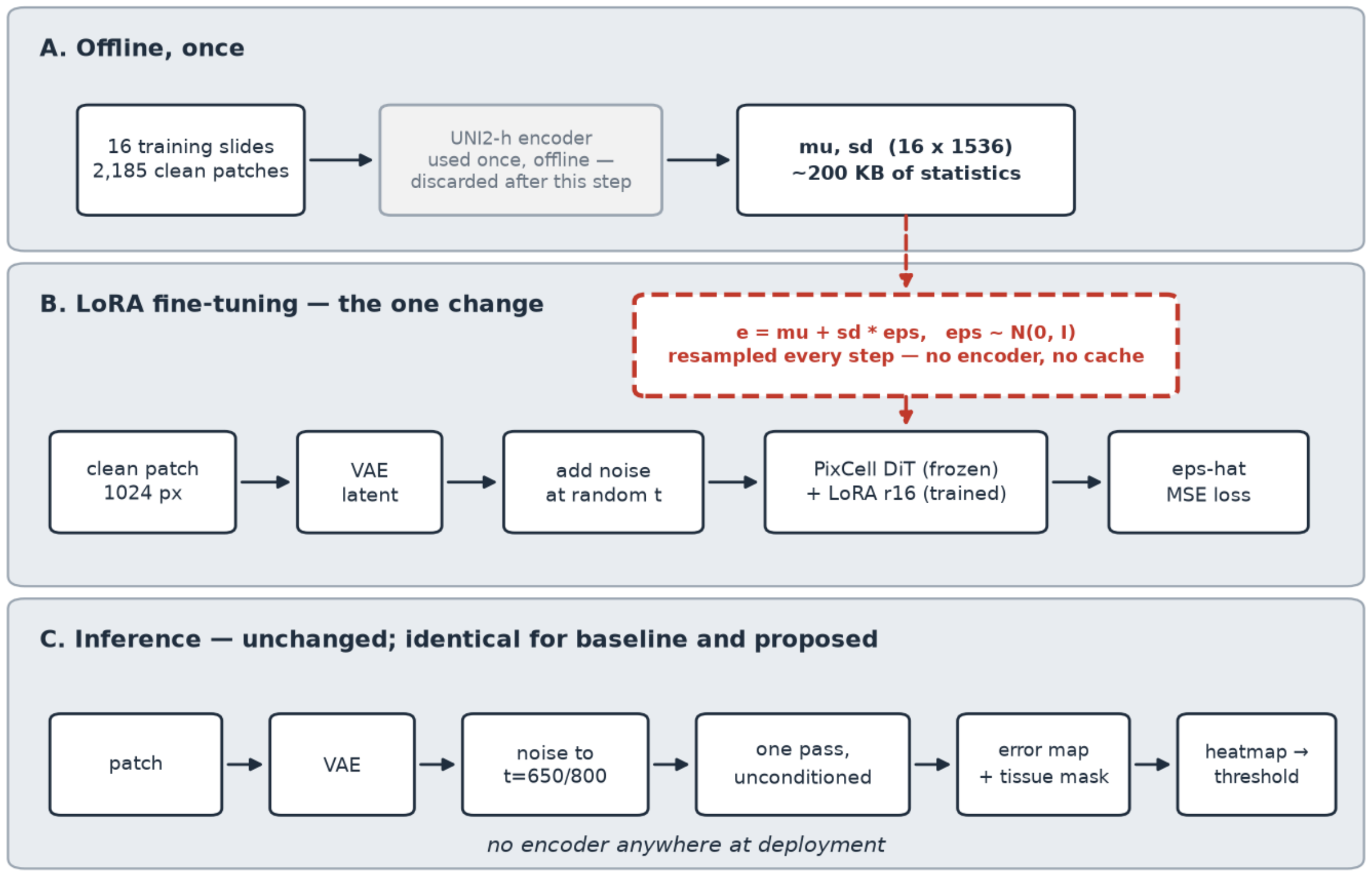

Figure 1. The proposed method as one change to a standard pipeline. (A) Offline, once: UNI2-h encodes the 2,185 clean training patches; only the per-position first and second moments (mu, sd — ~200 KB) are kept, and the encoder is discarded. (B) LoRA fine-tuning of the frozen PixCell DiT, identical to the baseline recipe except for the dashed red element: at every step the conditioning input is a fresh synthetic draw e = mu + sd * eps. (C) Inference is byte-identical for baseline and proposed models — one unconditioned pass at a fixed timestep, error map, tissue masking, percentile recalibration, thresholds carried from training slides. No pathology encoder appears anywhere at deployment.

## 4. Mechanism: what the conditioning benefit does — and does not — require

### 4.1 The final scheme, stated first

The complete mechanism is this: during LoRA fine-tuning, the cross-attention conditioning input of the diffusion transformer receives, at every training step and for every patch, a fresh random draw

```
e = mu + sd * eps,    eps ~ N(0, I),
```

where mu and sd are the per-token-position mean and standard deviation of UNI2-h embeddings computed once over the 2,185 clean training patches — 2 x 16 x 1536 floating-point values, roughly 200 KB. Inference is unchanged in every byte: scoring uses the null embedding, exactly as the unconditioned baseline. No image encoder is invoked at any stage;

the statistics themselves are the only trace the foundation model leaves in the pipeline. Trained this way, the detector's clean/artifact separation at the patch level rises from Cohen's d = 1.34 to 1.76 (t = 650), with seed-42/seed-7 replicates essentially coincident (1.762 and 1.755 — 0.007 apart in Cohen's d, against a seed-noise floor of 0.126) — the tightest pair among the twelve conditioned trainings of this family — and the slide-level and external gains reported in Section 6.

The remainder of this section justifies each simplification in that recipe by elimination: four properties one might assume essential — content, provenance, intensity, identity — are shown individually unnecessary, and one control shows what *is* necessary: the pathway.

### 4.2 Content is unnecessary

Conditioning each patch on its own UNI2-h embedding ("self-conditioning") improves d from 1.34 to 1.59–1.67 across three training seeds. Randomly permuting the patch-embedding correspondence — every patch conditioned on some *other* patch's embedding — does not reduce the benefit: d = 1.79, at or above every correctly-paired seed. Whatever the conditioning contributes, it is not information about the patch being denoised.

### 4.3 Provenance is unnecessary

If content is irrelevant, are real embeddings needed at all? Replacing the cache with samples from a per-token-position diagonal Gaussian matched to the real embeddings' first two moments yields d = 1.78 / 1.71 (two seeds), indistinguishable from the shuffled-real arm. An independent redraw of the synthetic cache (sampling seed 1) lands at 1.77 — a three-run family (d = 1.783/1.714/1.775) whose range, 0.069 in Cohen's d, is half the measured seed noise. The encoder is needed once, offline, to estimate mu and sd; no real embedding ever enters training.

### 4.4 Intensity is not a tunable

Scaling the synthetic standard deviation across an 8x range (x0.5, x1, x2, x4) moves d by less than the seed noise: 1.81, 1.71–1.78, 1.73, 1.82. The x4 point is notable: its token norms (~58) sit 3.6x outside anything the pretrained cross-attention encountered (~16), yet the benefit is fully preserved. The mechanism has no sensitive hyperparameter — a deployment property we consider stronger than a sharp optimum would be.

### 4.5 Identity is unnecessary

All caches above assign each patch a fixed random embedding — a consistent "pseudo-identity" across training. Redrawing eps at every step (the final scheme) removes even that: d = 1.76 / 1.755 (two seeds), inside the family band. The model does not need to associate patches with stable keys; uncorrelated per-step perturbation suffices. This eliminates memorization-based accounts and places the effect in the noise-injection-as-regularization family [12] (formalized for our setting in Section 4.8): perturbing an input during training penalizes the network's sensitivity to that input, here tightening the learned clean manifold against the identity shortcut.

### 4.6 The pathway is necessary

Is any training perturbation equivalent? A control replaces conditioning noise with LoRA dropout (rate 0.1, no conditioning of any kind): d = 1.23 — an excess of −0.11 over baseline, null within seed noise (and with a distinct low-t fingerprint, reported descriptively in the released record). Generic weight-space stochasticity does not reproduce the effect at this probe; the regularization acts through the conditioning/cross-attention input specifically.

### 4.7 Summary of the ablation chain

Table 1 collects the chain: one row per ablated property, with the patch-level separation each variant retains and the verdict it supports. The full pre-registration, reading rule, and dated verdict of every row live in the released record (entries E1ζ–E1ι, Section 8).

| ablated property | arm | d (t650) | verdict |
|---|---|---|---|
| (baseline) | unconditioned | 1.34 | — |
| — | self-conditioned, 3 seeds | 1.59–1.67 | benefit exists |
| content | shuffled pairing | 1.79 | unnecessary |
| provenance | synthetic Gaussian, 2 seeds + redraw | 1.71–1.78 | unnecessary |
| intensity | variance x0.5–x4 | 1.71–1.82 | no tunable |
| identity | fresh per-step (final), 2 seeds | 1.755–1.76 | unnecessary |

| pathway | LoRA dropout control | 1.23 | **necessary** |
|---|---|---|---|

Table 1. The ablation chain in one view: each row removes one property of the conditioning signal and reports the resulting patch-level separation at the pre-registered t = 650 axis. Every content-free variant preserves the benefit; only the pathway control (LoRA dropout in place of conditioning noise) loses it. Ranges span the seeds/redraws of each arm.

### 4.8 A first-order account

The classical noise-injection result [12] transfers to our setting directly. Write the fine-tuning objective with a noised conditioning input e = mu + s * eta, eta ~ N(0, I) (elementwise product; s the applied per-dimension standard deviations), and expand the denoiser to first order around the mean embedding:

```
eps_theta(z_t, t, e) ~= eps_theta(z_t, t, mu) + J (s * eta),
```

where J is the Jacobian of the predicted noise with respect to the conditioning input at mu. In expectation over eta the linear cross term vanishes (eta is zero-mean), leaving

```
E[loss] ~= loss_at_mu + sum_ij s_j^2 J_ij^2,
```

i.e. the base objective plus a Tikhonov penalty on the conditioning-input Jacobian, weighted by the injected per-dimension variances. (The expansion also yields a residual-times-curvature term of the same order in s; Bishop's argument removes it at the solution: the optimal least-squares predictor is the conditional mean of the target, and our target eps does not depend on the conditioning input at all, so the residual's conditional average — and with it that term — vanishes at the optimum. This is precisely Bishop's reduction, for sum-of-squares error, to a first-derivative Tikhonov form.) Fine-tuning under fresh conditioning noise is therefore, to first order, fine-tuning with an explicit sensitivity penalty on the cross-attention pathway — pressure toward adapters that rely less on the conditioning signal and encode cleaner unconditional structure.

This single expression anticipates most of the ablation table. Only the first two moments of the conditioning distribution enter, so the *content* and *provenance* of the embeddings cannot matter (Sections 4.2–4.3), and fixed versus fresh draws coincide in expectation (Section 4.5). The penalty acts on the input of one specific pathway, so generic weight-space perturbation is not equivalent (Section 4.6). One observation the first-order account does not by itself explain is the intensity plateau: the penalty coefficient scales with the squared multiplier, yet the measured effect is flat across 8x (Section 4.4) — indicating the benefit is not proportional to penalty magnitude in this regime, and marking the boundary where the first-order story ends (the classical equivalence is itself a small-noise result; the x4 arm sits well outside that limit). We offer this account as the parsimonious frame consistent with all seven ablations, not as proof (Section 7.4).

## 5. Evaluation protocol

The external effects this paper reports (ΔF1 of +0.007 to +0.013) are comparable to — or smaller than — two biases we measured in the same pipeline: the +0.016 F1 optimism of selecting thresholds on the evaluation slides themselves (the field's common practice), and the 0.011 F1 standard deviation between trainings that differ only in random seed. At this scale, an evaluation that does not measure and remove its own biases cannot distinguish its findings from noise. The protocol below exists so that every reported gain is legible against measured noise and priced selection effects; we consider it a contribution alongside the method.

### 5.1 Measured noise floors

Three baseline trainings differing only in seed set the patch-level noise floor at s_seed = 0.126 (t = 650 Cohen's d) and the slide-level floor at s_slide = 0.0112 (pooled F1). Every comparison in this paper is stated in units of these floors, and no single-run difference below 2 x s_seed was ever interpreted as an effect. Replicates were run for every load-bearing claim (Section 4's table lists seeds per arm).

### 5.2 Honest slide-level selection

Slide-level F1 requires choosing a heatmap threshold configuration. Choosing it on the evaluation slides themselves — the field's common "grid-max" practice — inflates F1 by +0.016 on our development set, an optimism we measured by running both protocols side by side. All slide-level numbers in this paper use leave-one-slide-out selection instead: each held-out slide is scored under the configuration chosen on the remaining slides only, and the grid-max figure is reported once, separately, as the price of the shortcut. Development-set comparative power is honestly limited — 24 slides with a Kish effective n of 6.2 — which is precisely why claims graduate to external confirmation.

### 5.3 Pre-registered external confirmation

External evaluation on GrandQC MPP10 followed a frozen two-look budget, declared before the first look: one primary endpoint per look (pooled ΔF1 of a named model pair), label-free percentile recalibration at percentiles fixed in advance (65, 75) on each model's own pooled score distribution, dev-frozen postprocessing, and a stratified paired case-level bootstrap (20,000 replicates, fixed seed). The second look carried a Bonferroni-style correction (97.5% CI) declared at reservation time, and its point-estimate prediction band was recorded before the run. Hypotheses arising *from* external findings (e.g. the fold transfer collapse of Section 6.3) are excluded from confirmation on the same set — they require new data, a discipline we adopted to keep the external set from silently becoming a development set. Determinism checks (bit-identical recalibration thresholds from independently produced canvases) gate every look.

### 5.4 Provenance

Each run in this study writes a record of its git commit, arguments, and environment beside its outputs, and each experiment's pre-registration, reading rule, and dated result live in a version-controlled lab notebook released with the code (Section 8). A reader can therefore check, for any reported number, when its protocol was frozen relative to its result — including the places where execution deviated from plan, which remain in the record as they occurred.

## 6. Results

### 6.1 Patch level: a twelve-training family

Table 2 collects every training of the conditioning family under the standing diagnostic (unconditioned scoring, t in {500, 650, 800} — t = 650 the pre-registered reading axis, t =

800 the deployment operating point; the t = 500 column is retained in the released per-run reports but not tabulated here — fixed noise seed). At t = 650 the unconditioned baseline separates clean from artifact patches at d = 1.34; every conditioned variant — self, shuffled, synthetic (three draws/seeds), variance-scaled (x0.5–x4), and fresh per-step (two seeds) — lies in the band d = 1.59–1.82 (Figure 2), an excess of +0.25 to +0.48 against a measured training-seed noise of s_seed = 0.126. No variant falls back to baseline; no hyperparameter of the perturbation (content, provenance, intensity, identity — Section 4) moves the result outside the band. At t = 800 every variant again sits above baseline (family 2.50–2.79 vs 2.42), though individual excesses there are not all beyond the t = 800 seed noise (s = 0.186); t = 650 is the pre-registered reading axis throughout. Air-bubble and out-of-focus per-type separations are stable across the family (1.35–1.55 and 1.99–2.16 respectively at t = 650), indicating the regularizer widens the margin globally rather than re-ranking artifact types.

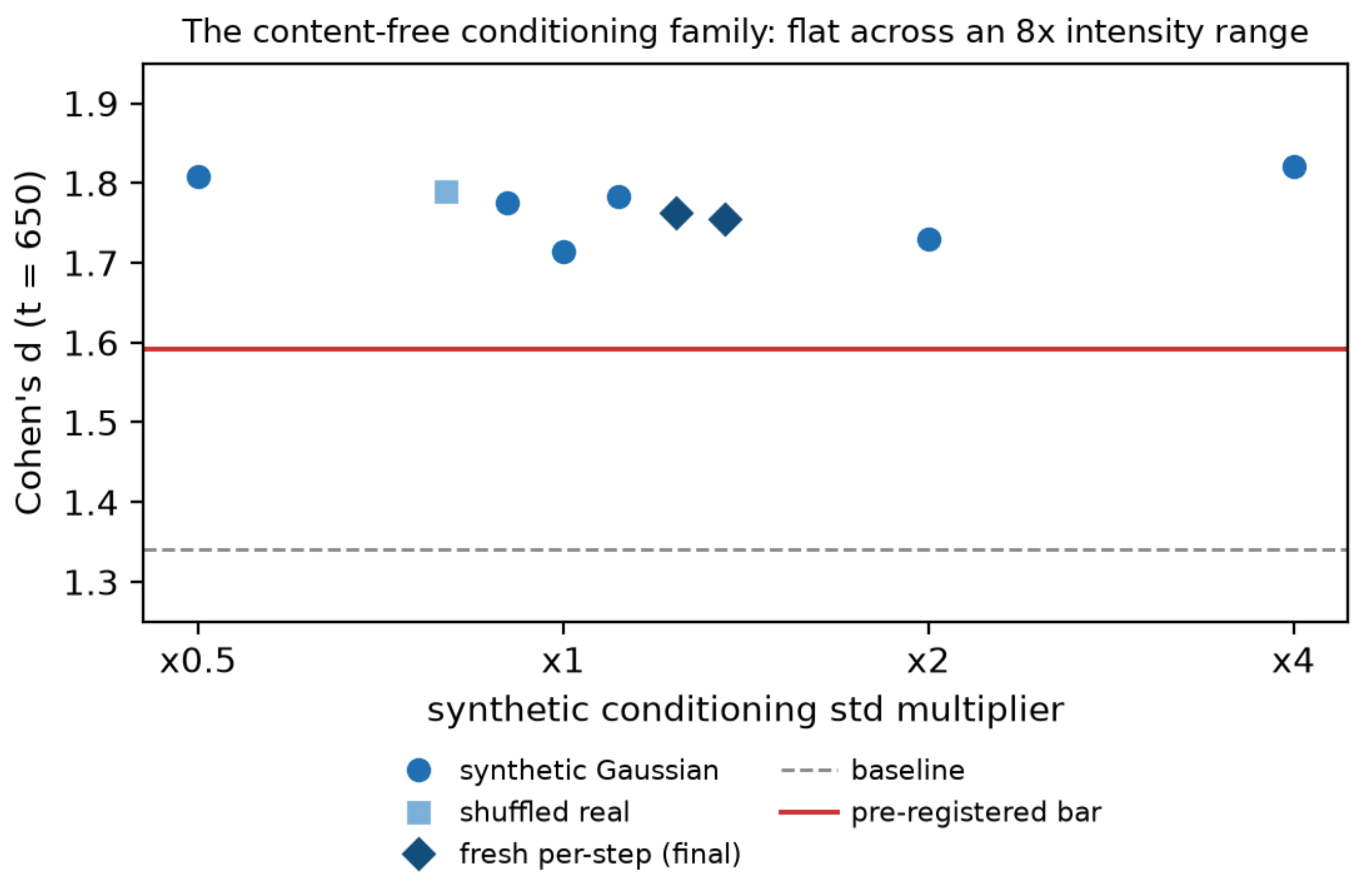

Figure 2. Patch-level separation (Cohen's d at the pre-registered t = 650 axis) for the nine content-free conditioned trainings, against the baseline (dashed, d = 1.34) and the pre-registered effect bar (red: baseline + 2 x s_seed = 1.59, where s_seed = 0.126 is the seed-noise floor of Section 3.5). Circles: the six synthetic-Gaussian trainings of the variance ladder (x0.5 to x4; the three x1 points are horizontally jittered for visibility). Square: shuffled real embeddings (content ablation); diamonds: fresh per-step resampling, the final configuration (two

seeds) — both plotted near x1, slightly offset horizontally, since they share the x1 statistics. All nine trainings clear the bar; separation is flat across the 8x intensity range. The LoRA-dropout control (d = 1.23, Section 4.6) is not part of the conditioning family and is reported in the mechanism table.

| arm | group | d (t650) | d (t800) |
|---|---|---|---|
| baseline | baseline | 1.340 | 2.416 |
| self s42 | self | 1.665 | 2.583 |
| self s7 | self | 1.599 | 2.791 |
| self s123 | self | 1.590 | 2.623 |
| shuffled | contentfree | 1.789 | 2.528 |
| synth s42 | contentfree | 1.783 | 2.543 |
| synth s7 | contentfree | 1.714 | 2.758 |
| synth redraw | contentfree | 1.775 | 2.560 |
| var x0.5 | ladder | 1.809 | 2.561 |
| var x2 | ladder | 1.731 | 2.510 |
| var x4 | ladder | 1.822 | 2.503 |
| fresh s42 | final | 1.762 | 2.550 |
| fresh s7 | final | 1.755 | 2.761 |
| dropout ctrl | control | 1.231 | 2.406 |

Table 2. The twelve conditioned trainings and the baseline: patch-level Cohen's d at t = 650 (pre-registered axis) and t = 800, one row per training, grouped by family (self / shuffled / synthetic + variance ladder / fresh) with the dropout control listed separately.

### 6.2 Slide level: honest leave-one-slide-out

Slide-level F1 is evaluated with the protocol of Section 5 (Table 3a): per-slide threshold menus, leave-one-slide-out selection, no test-slide information in any choice, and the selection optimism of the alternative grid-max protocol priced separately (+0.016 F1; Section 5). The unconditioned baseline reaches pooled F1 = 0.6567 with a measured slide-level seed noise of s_slide = 0.0112, setting a pre-registered bar of 0.6679. The final configuration clears it in both seeds: **0.6716 and 0.6714** (+0.0149/+0.0147; seed spread 0.0002, the tightest pair recorded at this level). The earlier self-conditioned variant had cleared the same bar in two of three seeds (0.6764/0.6661/0.6730) — the final configuration is not only simpler but more stable.

### 6.3 External confirmation: two pre-registered endpoints

Both external claims were declared, frozen, and corrected for multiplicity before any external number was read (protocol in Section 5.3; Table 3b, Figure 3). On the 281-case GrandQC

MPP10 set (33,361 blocks; 134.4M valid cells per model; label-free percentile recalibration fixed in advance):

- **Look one (self-conditioned vs baseline):** pooled ΔF1 = **+0.0073**, stratified paired case bootstrap 95% CI (+0.0042, +0.0110) → confirmed. - **Look two (final configuration vs baseline; 97.5% CI under the declared two-look correction):** pooled ΔF1 = **+0.0129**, CI (+0.0068, +0.0201) → confirmed at the stricter level. Before this run, the two-look protocol had us record a forecast of where the point estimate would land (+0.005 to +0.010, Section 5.3); the observed +0.0129 came in above that band — the effect was larger than our own pre-declared expectation.

Determinism checks passed in both looks (identical baseline recalibration thresholds from independently produced and reused canvases: 0.26978/0.31806). In both looks the gain concentrates in out-of-focus (sensitivity 0.653 → 0.672 and 0.653 → 0.698 respectively); fold, air-bubble, and dark-spot sensitivities are ties. Fold sensitivity collapses from 0.47 (development) to 0.04–0.05 (external) in *all* models — a transfer failure of the operating point, not of the proposed training, and a finding we report as design guidance (Section 7).

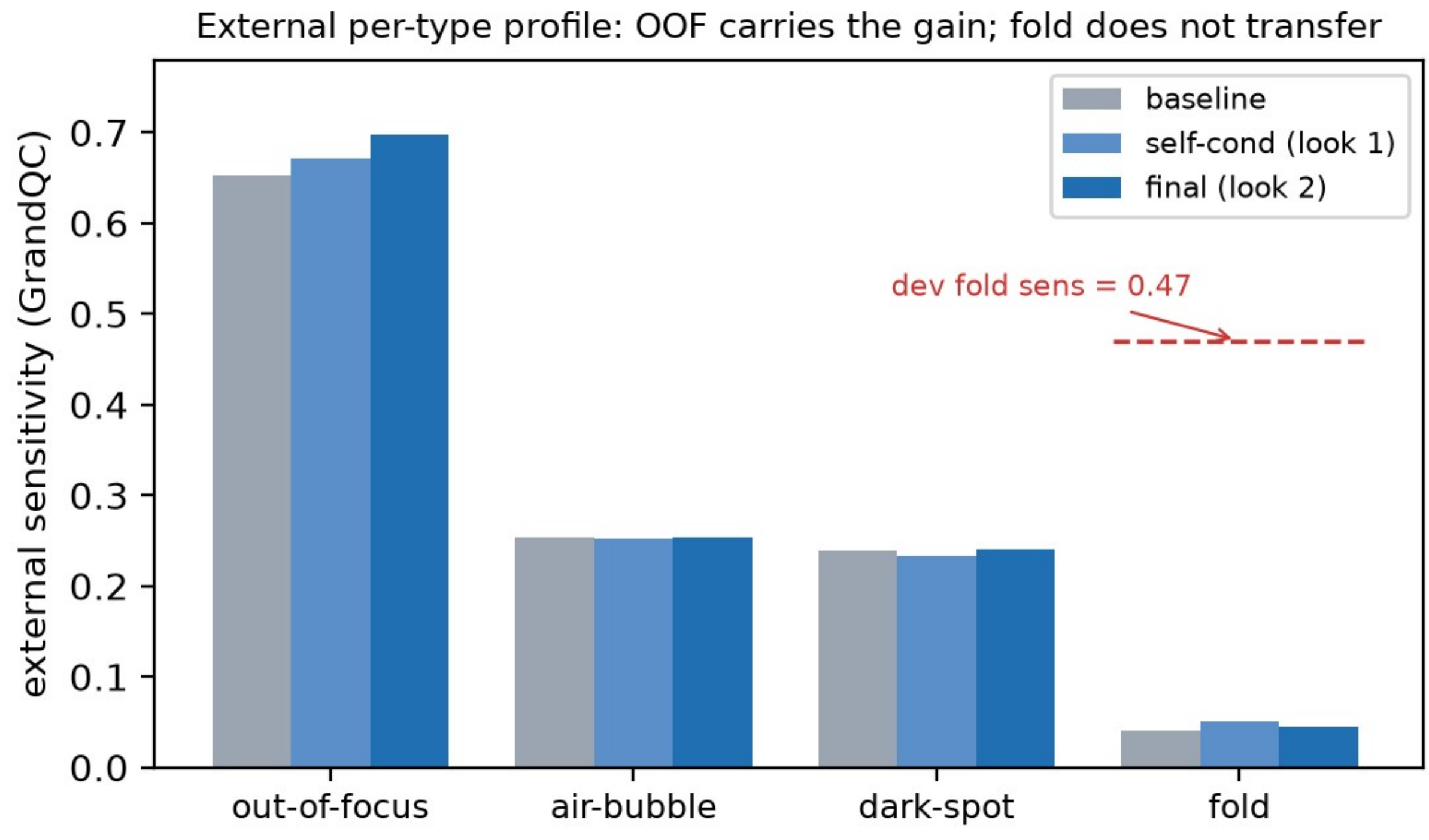

Figure 3. External per-type sensitivity on the GrandQC MPP10 cohort (281 cases) for the baseline, the look-1 model (self-conditioning), and the look-2 final model (fresh synthetic conditioning), at matched per-model

operating points. The confirmed gains concentrate in out-of-focus; air-bubble and dark-spot are ties. Fold sensitivity is near zero for every model including the baseline — against a development-set fold sensitivity of 0.47 (dashed red reference; note the development and GrandQC fold populations differ in annotation taxonomy and morphology, Section 7.2) — a symmetric transfer failure that bounds single-timestep operation, not a deficit of the proposed training.

**Table 3a.**

| model | pooled F1 | vs bar |
|---|---|---|
| baseline (E1δ record) | 0.6567 | — |
| self-cond s42/s7/s123 (E1ε record) | 0.6764 / 0.6661 / 0.6730 | 2/3 clear |
| final s42 | 0.6716 | clears |
| final s7 | 0.6714 | clears |

**Table 3b.**

| look | pair | ΔF1 | CI | level | verdict |
|---|---|---|---|---|---|
| 1 | self-cond − baseline | +0.0073 | (+0.0042, +0.0110) | 95% | CONFIRMED |
| 2 | final − baseline | +0.0129 | (+0.0068, +0.0201) | 97.5% | CONFIRMED |

Table 3. (a) Slide-level honest LOOCV on the 24-slide development test set: pooled F1 against the pre-declared bar (baseline + 1 x s_slide = 0.6679). (b) The two pre-registered external endpoints on GrandQC: ΔF1 with bootstrap CI (95% for look 1, 97.5% for look 2 under the two-look alpha budget), both CONFIRMED.

## 6.4 Why diffusion: the density-family baseline

Three feature-density scorers over the same UNI2-h embeddings (fit on the 2,185 clean training patches; scored on the 1,427-patch validation pool) answer the natural "why not model the features directly" question (Table 4):

| scorer | pooled d | AUROC | d bubble | d OOF |
|---|---|---|---|---|
| kNN (k=5, cosine) | 1.29 | 0.770 | 2.42 | 0.34 |
| Mahalanobis (Ledoit–Wolf) | 0.39 | 0.608 | 0.79 | −0.03 |
| GMM (8, diag) | 0.80 | 0.760 | 0.77 | 0.81 |

Table 4. Feature-density baselines (kNN, Mahalanobis, GMM over UNI2-h embeddings of the same validation pool) against the diffusion detector: pooled Cohen's d and AUROC, plus per-type d for the two classes where the families diverge most — the density family sees air-bubbles and misses defocus; the diffusion detector shows the reverse ordering (Section 6.4).

Every density scorer trails the diffusion detector pooled (best 1.29 vs 1.76/2.55) — but the per-type pattern is the finding: the two families rank the artifact types in opposite order. At the pre-registered t = 650 axis, kNN separates air-bubbles far better than defocus (2.42 vs 0.34) while the diffusion detector shows the reverse ordering (1.4–1.6 vs 2.1) — and the diffusion detector's externally confirmed gains concentrate in out-of-focus, precisely the class the feature family misses. A consistent reading: foundation features are trained toward photometric invariance, so defocus barely moves them; reconstruction error lives where blur *is* the signal. The two families are complementary by construction — we return to this in Section 7 — and, unlike the proposed method, every density scorer requires the foundation model at inference.

## 7. Discussion

### 7.1 What the effect is, and how large

At patch level the proposed regularizer is a substantial effect: +0.25–0.48 Cohen's d, a 20–33% relative widening of the clean/artifact margin, replicated across twelve trainings. By the time it reaches an external, frozen-protocol F1, it is small: +0.0073 and +0.0129. We frame this attenuation as information, not embarrassment: the chain from patch separation to deployed F1 discounts by roughly 5x in our measurements (threshold selection, spatial aggregation, per-type mixture), and the external gain concentrates where the patch-level gain lives (out-of-focus). A method whose development-set promise survives two pre-registered external looks at a known discount is, we argue, better characterized than one reporting a larger but unpriced number. The practical reading is honest: for a deployment dominated by out-of-focus detection, the regularizer buys a real margin for nothing; it does not rescue artifact types the base detector already misses.

### 7.2 Transfer is the open problem — and it is an operating-point problem

The starkest external finding is not about the proposed method: fold sensitivity collapses from 0.47 (development) to 0.04–0.05 (external) in every model, baseline included. The GrandQC fold population — smaller, thinner structures than our development folds — is nearly invisible at the t = 800 operating point. Because the collapse is symmetric it does not qualify the confirmed endpoints, but it bounds what any single fixed-timestep detector can claim across sites. Our development-set evidence (inverted per-type timestep profiles; the failure of global fusion schemes) points to per-type operating points as the path — a hypothesis that, having been motivated by the external set, must confirm on new data by our own protocol.

### 7.3 Data curation dominates architecture

Two development-set results discipline the field's default instincts. Scaling clean training data 4x while relaxing its curation gates — the patch-admission criteria of Section 3.2 (a minimum tissue fraction, plus exclusion of annotated artifact regions), progressively loosened so that background-heavier material qualifies as "clean" — *inverted* air-bubble separation ($d = -1.31$ at the loosest setting): more data, worse detector. The vulnerability is a documented one for likelihood-based generative detectors, which can rate low-complexity inputs, or inputs dominated by background statistics the model represents well, as confidently normal [21,22]; it appears here in reconstruction form, and our ablation adds the switch: the failure is not intrinsic but curation-induced, since the relaxed gate admitted background-like clean material whose statistics air bubbles share. The strict gate acts, in effect, as the curation-level counterpart of the scoring-time background corrections proposed in that line of work. And removing an artifact class from contrastive training data had no effect on detecting that class. Both point the same way: what the clean manifold *excludes* matters more than how much it includes, and curation choices deserve the ablation budget that architecture usually receives.

### 7.4 Limitations

One backbone (PixCell-1024), one stain (H&E), one magnification (10x), one adapter family (LoRA r16); the confirmed gains concentrate in one artifact type; the external set contributes no clean-slide stratum, so specificity claims rest on development data; and the mechanism account, while behaviorally complete (Section 4), remains first-order — we offer the noise-injection reading as the parsimonious theory consistent with all seven ablations, not as proof.

### 7.5 Future work

Several directions follow naturally. Per-type or adaptive timestep policies could address the transfer problem of Section 7.2, and a few annotated examples per artifact type may suffice to calibrate them. The inverse specialization of Section 6.4 invites a complementary ensemble of reconstruction- and feature-based scores; we left it outside this paper's scope since it reintroduces the encoder at inference, but the near-orthogonal error profiles make it a natural next step where that cost is acceptable. On the mechanism side, richer weight-space perturbation structures than our single dropout probe could map where the conditioning pathway's specialness ends. We would evaluate any of these under the same discipline applied here — pre-registered endpoints on data not previously consulted.

## 8. Reproducibility statement

All code, the trained-adapter checkpoints behind every reported number, and the complete lab record — pre-registrations, frozen reading rules, dated results, and documented deviations — are released at https://github.com/kmouts/condnoise-histoqc (archived: doi 10.5281/zenodo.22702198). Each run writes a provenance sidecar (git commit, full arguments, environment) beside its outputs; the deposit is rebuilt by a single script whose integrity checks are part of the release. The external evaluation uses the public GrandQC MPP10 set; recalibration is label-free and fully specified, so both external endpoints are reproducible from the released checkpoints without access to our development data. The ~200 KB of embedding statistics sufficient to train the proposed configuration ship with the code.

## Declarations

**CRediT authorship contribution statement.** Konstantinos Moutselos: Conceptualization, Methodology, Software, Validation, Formal analysis, Investigation, Data curation, Visualization, Writing - original draft, Writing - review & editing. Ilias Maglogiannis: Conceptualization, Methodology, Supervision, Resources, Funding acquisition, Writing - review & editing.

**Declaration of competing interest.** The authors declare that they have no known competing financial interests or personal relationships that could have appeared to influence the work reported in this paper.

**Data availability.** All primary data are public: the whole-slide images from the TCGA data portal, the artifact annotations released by the AIRAQc authors, and the GrandQC external test cohort released by its authors under their own licence. The outputs generated in this study - per-run diagnostic reports, external evaluation outputs with their provenance records, the trained adapter checkpoints, and the version-controlled laboratory record - are deposited at Zenodo (code and laboratory record: doi 10.5281/zenodo.22702198; model outputs and evaluation data: doi 10.5281/zenodo.22702800).

**Code availability.** Training, scoring, and evaluation code, with the exact configuration of every reported run, is available at https://github.com/kmouts/condnoise-histoqc, archived at Zenodo (doi 10.5281/zenodo.22702198, release v1.0-submission).

**Funding.** This work was carried out within Project PRODIGY, Greek-Chinese Bilateral Research and Technology Cooperation, Grant No. DSEKR01-0021403, Programme Competitiveness 2021-2027, Ministry of Economy and Finance, Greece, ERDF, co-funded by the European Union.

**Ethics.** This study used publicly available, de-identified whole-slide images from The Cancer Genome Atlas, publicly released annotations derived from them, and the publicly released GrandQC evaluation cohort. No new human data were collected and no ethics approval was required.

**Acknowledgements.** The authors thank the AIRAQc and GrandQC groups for releasing their annotations and evaluation data publicly, without which this study would not have been possible. Computing resources were provided by the University of Piraeus, NVIDIA DGX GPU Cluster.

**Declaration of generative AI and AI-assisted technologies.** During the preparation of this work the authors used a large language model to help design and analyse the experiments, write evaluation code, and draft and revise the manuscript text. The authors reviewed and verified all code, results, and text, and take full responsibility for the content of the published article.